\documentclass[journal=jpc,manuscript=article]{achemso}
\usepackage{longtable}
\usepackage{multirow}
\usepackage{amsmath}
\usepackage{subfigure}
\usepackage[usenames,dvipsnames]{xcolor}
\usepackage{soul}
\usepackage{bm}

\usepackage{xr}
\usepackage{amssymb}
\usepackage{siunitx}
\usepackage{multicol} \usepackage{wrapfig} \usepackage{enumitem}
\usepackage{bm} \usepackage{outlines} 

\usepackage{url}

\usepackage{siunitx}\usepackage{booktabs}

\makeatletter

\SectionNumbersOn

\author{Silvan K\"aser} \affiliation[University of Basel]{Department
  of Chemistry, University of Basel, Klingelbergstrasse 80 , CH-4056
  Basel, Switzerland.}

\author{Markus Meuwly} \affiliation[University of Basel]{Department of
  Chemistry, University of Basel, Klingelbergstrasse 80 , CH-4056
  Basel, Switzerland.}  \email{m.meuwly@unibas.ch}

\title{Numerical Accuracy Matters: Applications of Machine Learned
  Potential Energy Surfaces}

\begin{document}
\date{\today}

\begin{abstract}
The role of numerical accuracy in training and evaluating neural
network-based potential energy surfaces is examined for different
experimental observables. For observables that require third- and
fourth-order derivatives of the total energy with respect to Cartesian
coordinates single-precision arithmetics as is typically used in
ML-based approaches is insufficient and leads to roughness of the
underlying PES as is explicitly demonstrated. Increasing the numerical
accuracy to double-precision yields a smooth PES with higher-order
derivatives that are numerically stable and yield meaningful
anharmonic frequencies and tunneling splitting as is demonstrated for
H$_2$CO and malonaldehyde. For molecular dynamics simulations, which
only require first-order derivatives, single-precision arithmetics
appears to be sufficient, though.
\end{abstract}

\pagebreak
\noindent
Traditional machine learning frameworks/models operate at 32-bit
(single precision) to reduce the computational cost for training and
inference and for memory efficiency. Much of this appears to be driven
by the notion that minimizing the loss function does not benefit from
increased accuracy. The performance of approximate, non-deterministic
computations in the context of neural network optimization and
inference was tested for deep and convolutional neural networks for
which the MNIST\cite{lecun:2010} (handwriting) and
CIFAR10\cite{hinton:2009} (image classification) datasets were
used.\cite{gupta:2015} For such applications it was found that the
noise-tolerance of neural networks combined with stochastic rounding
during training lead to results close to 32-bit computations. More
recently, even further reduced floating point precision, such as half
precision (16-bit), has been used.\cite{gupta:2015,wang2018training}
It should be noted that in these and more recent computations for
large language models typically only first derivatives are
required.\cite{peng:2023,micikevicius2022fp8} However, there may be
advantages in or even requirements for increased accuracy at the
inference step depending on the application considered, such as in the
natural sciences.\\

\noindent
One of the recent successful applications of machine learning in
molecular sciences is the construction of neural network-based
potential energy surfaces (PESs). The ability to conceive
full-dimensional PESs for molecules has revolutionised the field of
molecular
simulations\cite{manzhos2020neural,jiang2020high,unke2021machine,MM:2021,kaser2023neural}. This
is primarily due to their ease-of-use and versatility. Also, they have
shown to reproduce the input data quite accurately, with
representation errors well below chemical accuracy (1 kcal/mol). On
the other hand, compared with empirical energy functions, the number
of floating-point operations (FLOPS) for evaluating one energy and the
forces even for a small molecule can be large. For example, the neural
network underlying PhysNet\cite{MM.physnet:2019} contains of the order
of $10^6$ learnable parameters. This is based on a feature vector of
size $10^2$ (128) per atom which leads to $10^2 \times 10^2$ FLOPS for
each matrix-vector multiplication (MVM). There are $\sim 10$ MVMs per
module. For a network with 5 modules and for a molecule with 10 atoms
this yields $\sim 10^9$ FLOPS if application of the activation
functions is also accounted for. This estimate does not yet include
initialization and computation of the radial basis functions.\\
  
\noindent
The precision with which these numerical operations are carried out
during inference is decisive for the accuracy of one evaluation
(e.g. the energy or the forces) of the neural network. While for many
practical applications in atomistic simulations inaccuracies in the
total energy on the order of $\sim 10^{-5}$ kcal/mol remain without
consequences, depending on the observable of interest the accuracy
with which such calculations are carried out will be important. This
is considered in the present contribution for a perturbative treatment
of the molecular vibrations and tunneling splittings of medium-sized
molecules.\\

\noindent
The present contribution examines and quantifies how two NN-PESs based
on the PhysNet\cite{MM.physnet:2019} architecture perform depending on
whether 32-bit (NN$^{\rm F32}$) or 64-bit (NN$^{\rm F64}$) arithmetics
is used for training and evaluation. The systems considered are
(an)harmonic vibrations of formaldehyde (H$_2$CO) and tunneling
splittings for hydrogen transfer in malonaldehyde HO-(CH)$_3$O.\\

\noindent
The reference data ($E$, $\bm{F}$, $\bm{\mu}$) for the H$_2$CO PESs
was determined at the MP2/aug-cc-pVTZ level of
theory.\cite{mm.anharmonic:2021} The data set contains a total of 3601
structures for H$_2$CO obtained from normal mode
sampling.\cite{smith2017ani} For training, the 3601 structures were
split into training/validation/test sets according to 3061/360/180 and
the training was repeated twice on different splits of the data.
After initialization, the parameters of the NN are optimized by
minimizing a loss function using ADAM\cite{kingma2014adam} with a
learning rate of $10^{–3}$ and a batch size of 32 randomly chosen
structures. For malonaldehyde reference data at the MP2/aug-cc-pVTZ
(low-level, LL) and CCSD(T)/aug-cc-pVTZ (high-level, HL) levels of
theory are available from which PESs are
trained.\cite{mm.tlrpimalonaldehyde:2022} First, a LL representation
of the PES is obtained from $\sim 70000$ data points, which is then
transfer learned (TL) to the CCSD(T)/aug-cc-pVTZ level using the 862
configurations with corresponding $E$, $\bm{F}$,
$\bm{\mu}$.\cite{mm.tlrpimalonaldehyde:2022} For the NN$^{\rm F32}$
and NN$^{\rm F64}$ models everything except the floating-point
precision was kept unchanged.\\

\begin{table}
    \centering
    \begin{tabular}{rcccc}\toprule
                                       & \textbf{NN$_1^{\rm F32}$} & \textbf{NN$_2^{\rm F32}$} & \textbf{NN$_1^{\rm F64}$} &\textbf{NN$_2^{\rm F64}$} \\\midrule
    \textbf{MAE($E$)} kcal/mol         & 7.2E-3 & 1.3E-3 & 1.6E-5 & 1.0E-5\\
    \textbf{RMSE($E$)}kcal/mol         & 7.2E-3 & 1.3E-3 & 4.6E-5 & 1.5E-5\\
    \textbf{MAE($F$)} kcal/mol/\AA\/   & 1.4E-3 & 1.4E-3 & 5.0E-4 & 2.8E-4\\
    \textbf{RMSE($F$)}kcal/mol/\AA\/   & 3.7E-3 & 3.4E-3 & 3.0E-3 & 6.0E-4\\\bottomrule
    \end{tabular}
    \caption{Comparison of the test set errors for a total of 4
      PhysNet-based NN potentials for H$_2$CO. Two models operate at
      single- (NN$_i^{\rm F32}$) and two at double-precision
      (NN$_i^{\rm F64}$). The NN$_i^{\rm F64}$ models reach energy
      errors that are two orders of magnitude lower, while the effect
      on the forces is less pronounced but clearly visible.}
    \label{tab:out_of_sample_errors_h2co}
\end{table}

\noindent
The test set errors for all four PhysNet models for H$_2$CO are listed
in Table~\ref{tab:out_of_sample_errors_h2co}. NN$^{\rm F32}$ reaches
out of sample MAE($E$) of several $10^{-3}$ kcal/mol while for
NN$^{\rm F64}$ the energy errors are reduced by two orders of
magnitude. For the forces, NN$^{\rm F64}$ reduces the error to several
$10^{-4}$ kcal/mol/\AA\/ and lower, whereas NN$^{\rm F32}$ remains
consistently above $10^{-3}$ kcal/mol/\AA. For the following analysis,
the two models with lower out-of-sample errors are used
(\textit{i.e.}, NN$_2^{\rm F32}$ and NN$_2^{\rm F64}$).\\

\begin{table}
    \centering
    \begin{tabular}{cccccc}\toprule
        Harm. &  \textbf{NN$_2^{\rm F32}$} & \textbf{$|\Delta|$} &  \textbf{NN$_2^{\rm F64}$} & \textbf{$|\Delta|$} & \textbf{MP2 Ref.} \\\midrule
        1 & 1196.65 & 0.01 & 1196.66 & 0.00 & 1196.66 \\ 
        2 & 1266.70 & 0.03 & 1266.73 & 0.00 & 1266.73 \\ 
        3 & 1539.91 & 0.03 & 1539.96 & 0.02 & 1539.94 \\ 
        4 & 1752.47 & 0.10 & 1752.41 & 0.04 & 1752.37 \\ 
        5 & 2973.11 & 0.12 & 2973.12 & 0.11 & 2973.23 \\ 
        6 & 3047.10 & 0.18 & 3047.14 & 0.14 & 3047.28 \\ \midrule
        MAE &       & 0.08 &         & 0.05   &\\\bottomrule\bottomrule
        VPT2 &  \textbf{NN$_2^{\rm F32}$} & \textbf{$|\Delta|$} &  \textbf{NN$_2^{\rm F64}$} & \textbf{$|\Delta|$} & \textbf{MP2 Ref.} \\\midrule
        1 & 1179.94 & 0.29 & 1180.19 & 0.05 & 1180.23 \\ 
        2 & 1246.67 & 0.05 & 1246.65 & 0.06 & 1246.72 \\ 
        3 & 1507.47 & 0.48 & 1507.48 & 0.47 & 1507.95 \\ 
        4 & 1716.10 & 4.84 & 1718.81 & 2.13 & 1720.94 \\ 
        5 & 2823.11 & 3.55 & 2825.09 & 1.58 & 2826.67 \\ 
        6 & 2859.67 & 2.94 & 2860.78 & 1.83 & 2862.61 \\ \midrule
        MAE &       & 2.02 &         & 1.02   &\\\bottomrule
    \end{tabular}
    \caption{Harmonic frequencies determined by diagonalising the
      mass-weighted Hessian matrix and VPT2 frequencies as obtained
      from NN$_2^{\rm F32}$ and NN$_2^{\rm F64}$. All values are given
      in cm$^{-1}$.  The frequencies are compared to their \textit{ab
        initio} MP2/aug-cc-pVTZ level reference and the absolute
      difference $|\Delta|$ is shown. Both NN potentials yield
      harmonic frequencies of similar quality which do not exceed an
      absolute error of 0.2~cm$^{-1}$, although NN$_2^{\rm F64}$
      reaches lower errors on average. }
    \label{tab:harmonic_freq_h2co}
\end{table}

\noindent
Harmonic frequencies $\omega$, from diagonalising the mass-weighted
Hessian matrix, were determined with both NN$_2^{\rm F32}$ and
NN$_2^{\rm F64}$ and compared with \textit{ab initio} MP2/aug-cc-pVTZ
frequencies together with their absolute deviations in
Table~\ref{tab:harmonic_freq_h2co}. All computed harmonic frequencies
are within 0.2~cm$^{-1}$ from their reference and illustrate the high
quality of the NN-PESs around the minimum. While both floating-point
variants reach similar deviations with respect to the reference
harmonic frequencies with MAE($\omega$) of 0.08 and 0.05~cm$^{-1}$,
respectively, NN$_2^{\rm F64}$ is consistently more accurate.\\

\noindent
A widely used improvement of harmonic frequencies to include effects
of mechanical anharmonicity is to determine perturbative corrections
up to second order (VPT2).\cite{barone2005anharmonic} Such
calculations require third- and fourth-order derivatives of the energy
with respect to nuclear positions which need to be carried out
numerically. If carried out within the realm of {\it ab initio}
calculations this leads to prohibitively long computation times for
all but the smallest molecules and at sufficiently high levels of
quantum chemical theory. Recently, a viable alternative combining
ML-based PESs and TL demonstrated that VPT2 calculations at the
CCSD(T) level are possible for molecules the size of
CH$_3$CONH$_2$.\cite{mm.anharmonic:2021} The VPT2 frequencies using
NN$_2^{\rm F32}$ and NN$_2^{\rm F64}$ for H$_2$CO are given in
Table~\ref{tab:harmonic_freq_h2co} and are compared with {\it ab
  initio-}calculated values at the MP2/aug-cc-pVTZ level of theory.
While both model types perform reasonably well, the prediction error
of NN$_2^{\rm F32}$ is now often larger by a factor of two or more
compared to NN$_2^{\rm F64}$. The average difference for NN$_2^{\rm
  F32}$ and NN$_2^{\rm F64}$ is 2 cm$^{-1}$ and 1 cm$^{-1}$,
respectively.\\

\begin{figure}[ht]
\centering
\includegraphics[width=0.8\textwidth]{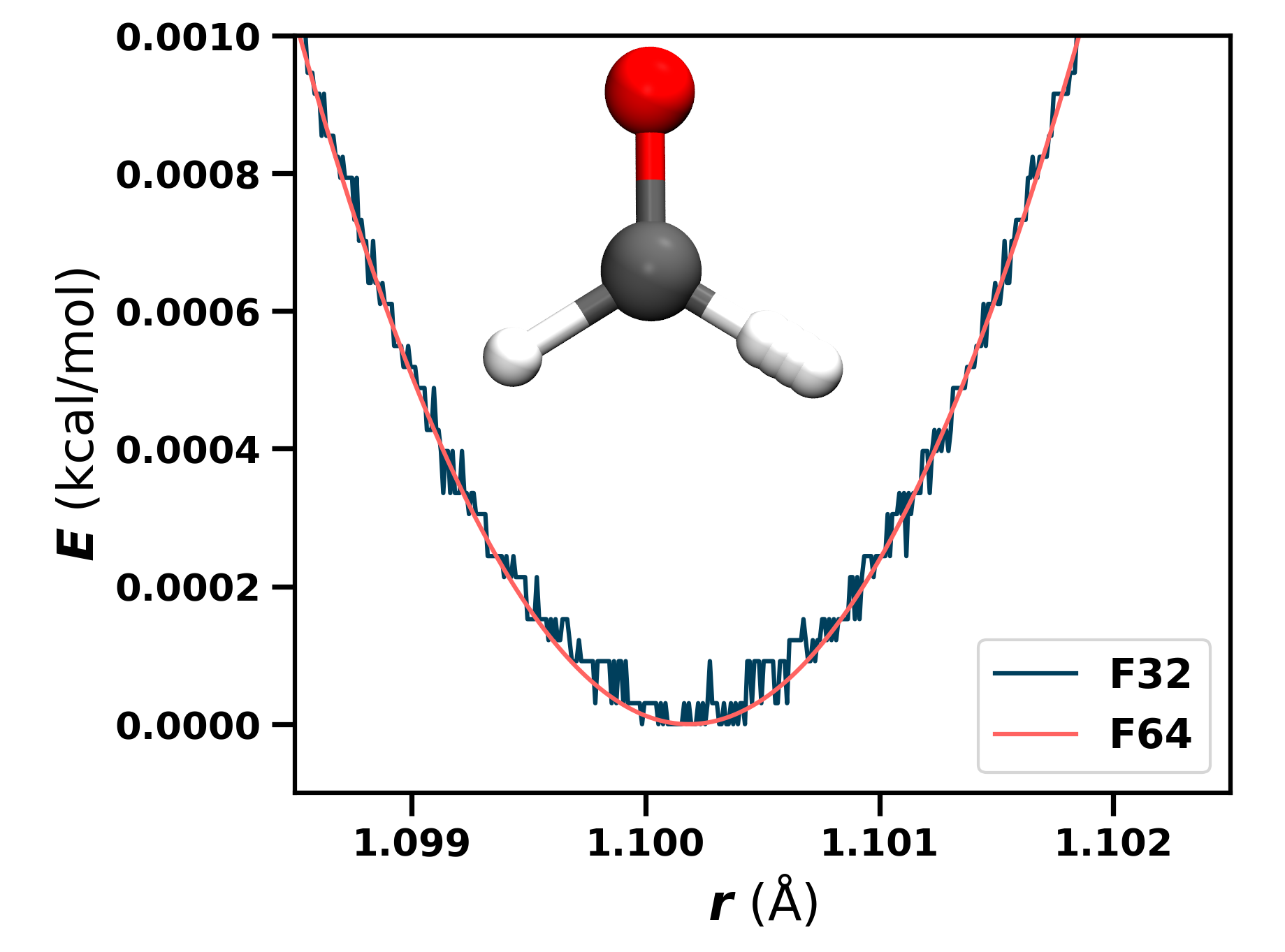}
\caption{High-resolution potential energy scan along a C--H bond of
  H$_2$CO with both NN$_2^{\rm F32}$ and NN$_2^{\rm F64}$. The scan
  has been recorded with a step size of $\Delta r = 0.00001$~\AA~ and
  only shows the region in the vicinity of the global minimum. From
  the potential energy scan, it is obvious that the NN$_2^{\rm F32}$
  PES has a higher ``roughness'' and its effect on computable
  observables remains to be clarified.}
\label{fig:energy_wiggles_f32_f64}
\end{figure}

\noindent
The most likely reason for the differences between NN$_2^{\rm F32}$
and NN$_2^{\rm F64}$ are inaccuracies in the numerical derivatives due
to small-scale variations of the underlying PES, \textit{i.e.}, due to
``roughness'' of the PES. In order to test this hypothesis, the two
PESs were scanned along a representative coordinate, which was the
C--H bond and using a small step size of $\Delta r
=10^{-5}$~\AA\/. The resulting energy scans around the minimum energy
structure are shown in Figure~\ref{fig:energy_wiggles_f32_f64}. It is
obvious that the NN$_2^{\rm F32}$ PES reveals small-scale energy
variations ($10^{-5}$~kcal/mol) whereas for NN$_2^{\rm F64}$ the PES
remains smooth viewed on the same energy scale. Given that the
harmonic frequencies are rather insensitive to whether NN$_2^{\rm
  F32}$ or NN$_2^{\rm F64}$ is used, but more pronounced differences
arise if higher-order derivatives are required as for VPT2
calculations, it is conceivable that ``roughness'' which is present
for NN$_2^{\rm F32}$ but not for NN$_2^{\rm F64}$ has a detrimental
effect on performance.\\

\noindent
One possibility to {\it quantify} roughness of a PES can be based on
concepts from atomic force
microscopy.\cite{antonio2012scale,bellitto2012atomic} Given a height
profile $z$ measured using an atomic force microscope, the roughness
can be characterised in terms of the root mean squared roughness given
by
\begin{align}
    R_{\rm rms} = \sqrt{\frac{1}{N} \sum_{i=1}^N |z_i^2|}. 
\end{align}
Here, $N$ corresponds to the number of data points or ``measurements''
and $z_i$ is the deviation of the measurement $i$ from a $n-$point
average. For the NN$_2^{\rm F32}$ and NN$_2^{\rm F64}$ PESs first a
smoothed PES was determined from a 5-point moving average. Next,
$R_{\rm rms}$ was calculated, which yields $R_{\rm rms} =
2\cdot10^{-5}$ and $R_{\rm rms} = 7\cdot10^{-8}$~kcal/mol for
NN$_2^{\rm F32}$ and NN$_2^{\rm F64}$, respectively.  For NN$_2^{\rm
  F32}$ this corresponds approximately to the magnitude of the
variations in Figure~\ref{fig:energy_wiggles_f32_f64}. It should be
noted that $R_{\rm rms}$ allows to quantify the roughness of a PES,
but its magnitude depends on the scanning step size and the window
used for computing the running average.\\

\noindent
For malonaldehyde, the NN$^{\rm F32}$ and NN$^{\rm F64}$ PESs were
used to obtain tunneling splittings from ring polymer instanton (RPI)
calculations.\cite{richardson:2018} Consistent with previous
work\cite{mm.tlrpimalonaldehyde:2022} that reported $\sim 25$
cm$^{-1}$, the NN$^{\rm F32}$ and NN$^{\rm F64}$ PESs yield tunneling
splittings of 25.3~cm$^{-1}$ and 24.2~cm$^{-1}$, compared with 21.6
cm$^{-1}$ from experiment.\cite{firth1991tunable} Depending on the
coordinate system used, fixed-node Diffusion Monte Carlo (DMC)
simulations on a PES based on near-basis-set-limit frozen-core CCSD(T)
electronic energies and represented as permutationally invariant
polynomials (PIPs) find $21.6 \pm 3$ cm$^{-1}$ and $22.6 \pm 3$
cm$^{-1}$.\cite{wang2008full} Using the same PES, RPI simulations
reported 25 cm$^{-1}$.\cite{jahr2020instanton} \\

\begin{table}[ht]
    \centering
    \begin{tabular}{rrcrc}
      \toprule
      step size &  & $\Delta_{\rm RPI}$ (cm$^{-1}$) & $c_{\rm CF}$ & $\Delta_{\rm RPI + PC}$ (cm$^{-1}$) \\
      \midrule
       0.001  & NN$^{\rm F32}$ & 25.3 & 8.91 & 225.4 \\
       &   NN$^{\rm F64}$  & 24.2 & 0.91 & 22.1 \\
       \midrule
       0.0005 & NN$^{\rm F32}$ & 25.3 & 21.74  & 550.0 \\
       &   NN$^{\rm F64}$ & 24.2 & 0.91  & 22.1 \\
       \bottomrule
    \end{tabular}
    \caption{Multiplicative correction factors $c_{\rm CF}$ to RPI
      tunneling splittings $\Delta_{\rm RPI}$, which are obtained with
      two different step sizes for numerical differentiation. The
      correction factors are obtained for both NN$^{\rm F32}$ and
      NN$^{\rm F64}$ and the perturbatively corrected tunneling
      splittings are calculated as $\Delta_{\rm RPIPC} = c_{\rm CF}
      \cdot\Delta_{\rm RPI}$. While $c_{\rm CF}$ is consistent across
      different steps sizes for NN$^{\rm F64}$, large differences are
      found for NN$^{\rm F32}$. The
      experimentally reported tunneling splitting for H-transfer is
      $\Delta_{\rm Exp.} = 21.6$
      cm$^{-1}$.\cite{firth1991tunable,baba1999detection}}
    \label{tab:wiggly_tunneling_splittingsl}
\end{table}

\noindent
Recently, a perturbative correction for RPI simulations was presented
which determines a multiplicative factor $c_{\rm CF}$ to account for
anharmonic effects and intermode coupling.\cite{lawrence:2023}
Computation of $c_{\rm CF}$ requires third- and fourth-order
derivatives of the total energy with respect to Cartesian coordinates
which need to be carried out numerically, similar to the situation for
VPT2 calculations. The corrections $c_{\rm CF}$ to the RPI splittings
for different step sizes in the numerical differentiation and for both
PESs are listed in Table~\ref{tab:wiggly_tunneling_splittingsl}. For
NN$^{\rm F64}$ the correction is $c_{\rm CF} = 0.91$ irrespective of
step size and yields a corrected tunneling splitting of 22.1 cm$^{-1}$
which is in excellent agreement with the experimental value of 21.6
cm$^{-1}$.\cite{firth1991tunable,baba1999detection} Contrary to that,
the ``correction'', which is expected to be $\sim 1$, ranges from 8.9
to 21.7 depending on the step size when using NN$^{\rm F32}$. Hence,
the roughness of NN$^{\rm F32}$ (see below) has a detrimental effect
for higher-order derivatives which renders the perturbative
corrections to RPI meaningless. Finally, using a semi-global
PES\cite{mizukami:2014}, very recent RPI+PC simulations also found
22.1 cm$^{-1}$.\cite{lawrence:2023}\\

\noindent
The roughness of the two PESs for malonaldehyde, NN$^{\rm F32}$ and
NN$^{\rm F64}$, was also assessed by computing $R_{\rm rms}$ along a
C--H bond. Maintaining the same parameters (\textit{i.e.}, step size
of $\Delta r =10^{-5}$~\AA~ for scanning the potential energy and a
window of $n=5$ for the running average) as for H$_2$CO, the roughness
for the NN$^{\rm F32}$ and NN$^{\rm F64}$ PESs for malonaldehyde are
$5\cdot10^{-5}$ and $8\cdot10^{-8}$~kcal/mol, respectively. Therefore,
for NN$^{\rm F32}$ the roughness increases by a factor of 2.5 compared
with NN$^{\rm F32}$ for H$_2$CO whereas that for NN$^{\rm F64}$
remains almost unchanged for the two molecules. It is interesting to
note that for NN$^{\rm F32}$ the increase in roughness and size in
going from H$_2$CO to C$_3$H$_4$O$_2$ is comparable.\\

\begin{figure}[ht]
\centering
\includegraphics[width=0.75\textwidth]{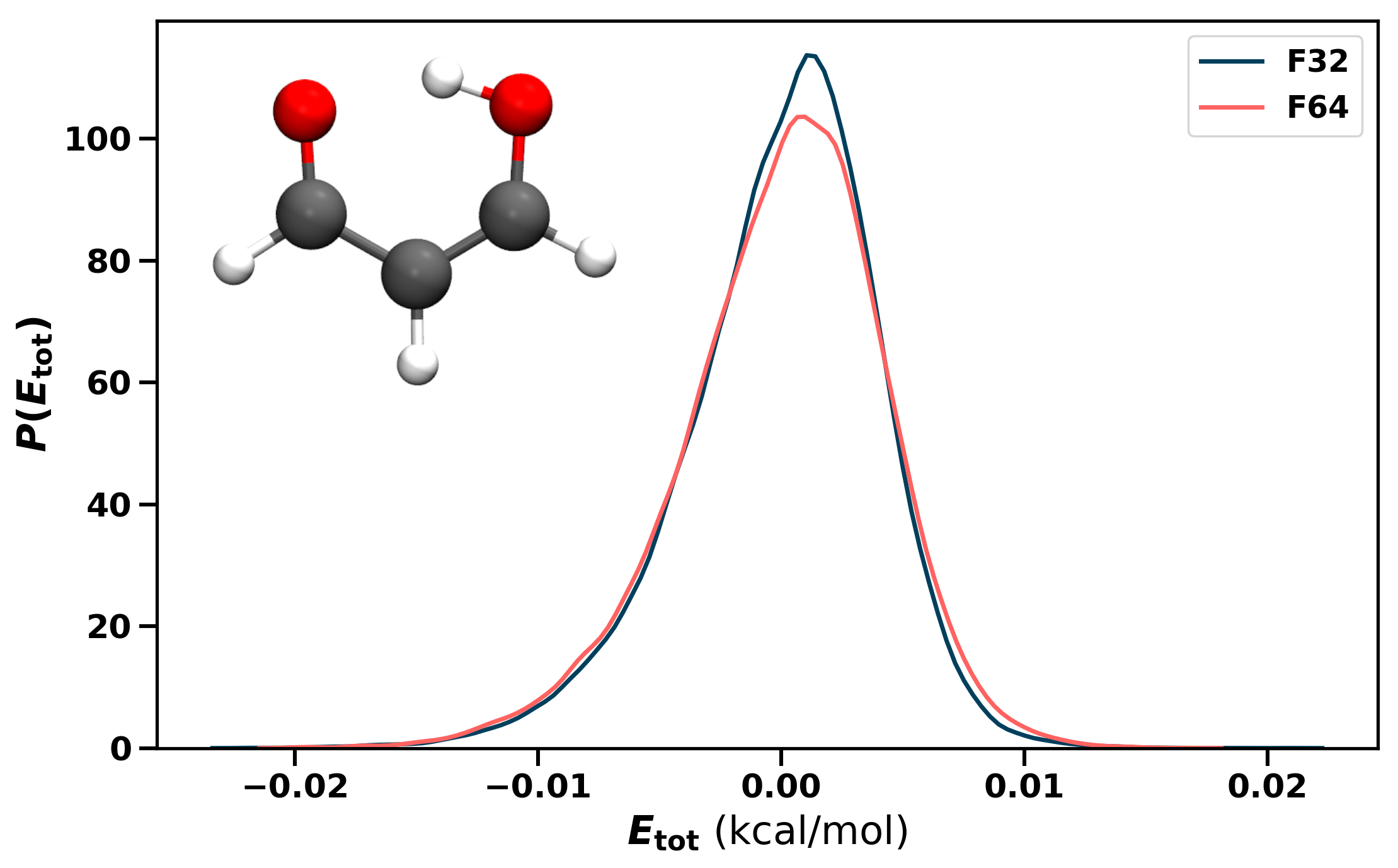}
\caption{Energy conservation for constant-$NVE$ simulations run at
  500~K with both NN$^{\rm F32}$ and NN$^{\rm F64}$.  A time step of
  0.25~fs and the Velocity Verlet integrator\cite{verlet1967computer}
  as implemented in ASE\cite{larsen2017atomic} were used. The total
  simulation time was 0.5~ns. Herein, the mean of $E_{\rm tot}$ is
  shifted to zero. The slight asymmetry in $P(E_{\rm tot})$ may be
    due to the fact that the dynamics is not fully equilibrated with
    respect to hydrogen transfer.}
\label{fig:energy_cons_f64_f34}
\end{figure}

\noindent
An important application of NN-based PESs is their use in atomistic
simulations, specifically molecular dynamics (MD)
simulations.\cite{mm.atmos:2020,chmiela:2023,mm.criegee:2021}
Therefore, energy conservation in gas-phase simulations was also
assessed in constant-$NVE$ simulations for malonaldehyde using the
NN$^{\rm F32}$ and NN$^{\rm F64}$ PESs. For this, the total energy
time series $E(t)$ was recorded from 0.5 ns simulations with a time
step of $\Delta t = 0.25$ fs using the Velocity Verlet
algorithm\cite{verlet1967computer} as implemented in
ASE.\cite{larsen2017atomic} Initially, the system was heated to 500
K. The histograms $P(E)$, reported in Figure
\ref{fig:energy_cons_f64_f34}, are near-Gaussian with a standard
deviation of $\sim 0.004$~kcal/mol, for both PESs NN$^{\rm F32}$ and
NN$^{\rm F64}$. Hence, the roughness in NN$^{\rm F32}$ does not have
deteriorating effects when used in $NVE$ simulations.\\

\noindent
The effects of single- and double-precision arithmetics was studied
previously for integrators of the equations of motion used in MD
simulations.\cite{shaw:2007} In a dedicated study the degree to which
total energy was not conserved in $NVE$ simulations was found to also
depend on the numerical accuracy used. Accumulation of errors led
simulations carried out with single-precision arithmetics using the
GROMACS MD code to severely violate the conservation of total energy
whereas changing to double-precision arithmetics conserved this
quantity on the nanosecond time scale for an extended
system. Interestingly, it was also possible to formulate a correction
scheme for integrating the equations of motion that conserved $E$ even
within single-precision arithmetics.\\

\noindent
The increased sensitivity of the gradients and higher-order
derivatives to the roughness of the PES is reminiscent of approximate
correlated \textit{ab initio} methods such as local
MP2.\cite{schutz1999low} Truncations by neglecting matrix elements
below a certain magnitude, in order to speed up the calculations, lead
to numerical precision issues for gradients and, even more acutely,
for second derivatives.\cite{wang2023local}\\

\noindent
Using double-precision arithmetics in the present case leads to long
training times (several weeks compared to days for H$_2$CO) and
evaluation times increase by 30 to 40~\%. It is concluded that
depending on the application, double-precision accuracy in training
and evaluating NN-based PESs is mandatory. On the other hand, for MD
studies, single-precision numerics is likely to be sufficient for
meaningful simulations.\\

\noindent
This work provides a critical assessment of the influence of
floating-point precision within the context of NN-based PESs. The
findings illustrate the exceptional accuracy of both NN$^{\rm F32}$
and NN$^{\rm F64}$, as is evident for the out-of-sample errors as well
as for the harmonic frequencies for H$_2$CO. As was demonstrated, if
higher-order numerical derivatives are required for the target
observable, \textit{e.g.}, anharmonic frequencies based on
perturbative approaches such as VPT2 or corrections to tunneling
splittings, ``roughness'' of the PES originating from reduced (single)
precision calculations deteriorate the performance (for VPT2
frequencies) or make computations entirely meaningless (correction to
tunneling splittings). On the other hand, ``roughness'' of a magnitude
found in the present work ($\sim 10^{-5}$ kcal/mol) appears to be
inconsequential for MD simulations which only require first
derivatives.\\

\section*{Acknowledgment}
Support by the Swiss National Science Foundation through grants
200021{\_}117810, 200021{\_}215088, the NCCR MUST (to MM), and the
University of Basel is acknowledged. This work was also partly
supported by the United State Department of the Air Force, which is
gratefully acknowledged (to MM). The authors thank
Prof. J. O. Richardson and Dr. J. E. Lawrence for sharing computer
codes to carry out the perturbative corrections and for discussions.

\section*{Data Availability Statement}
The PhysNet Codes are available from
\url{https://github.com/MMunibas/PhysNet}, the H$_2$CO data set is
accessible at \url{https://zenodo.org/records/4585449} and the MP2
level data for malonaldehyde can be found at
\url{https://github.com/MMunibas/beta-diketones}.\\

\bibliography{refs}
\end{document}